\documentstyle[11pt]{article}
\newcommand{\beq}{\begin{equation}}
\newcommand{\eeq}{\end{equation}}
\newcommand{\beqr}{\begin{eqnarray}}
\newcommand{\eeqr}{\end{eqnarray}}
\newcommand{\nn}{\nonumber}

\newcommand{\sech}{{\rm sech}}

\newcommand{\sss}{\vspace{.2in}}
\newcommand{\sms}{\vspace{.1in}}
\begin{document}
\sss
\begin{center}
{\Large {\bf Potentials with Two Shifted Sets of Equally Spaced Eigenvalues
and Their Calogero Spectrum }}
\end{center}
\vspace{.8in}
\begin{center}
{\large{\bf Asim Gangopadhyaya$^{a,}$\footnote{asim@uicws.phy.uic.edu, 
agangop@orion.it.luc.edu ~~FAX: (773) 508-3534}
and Uday P. Sukhatme$^{b,}$\footnote{sukhatme@uic.edu ~~FAX: (312) 996-9016}  }}
\end{center}
\vspace{.8in}
\noindent
a) \hspace*{.2in}
Department of Physics, Loyola University Chicago, Chicago, Illinois 60626
\\
b) \hspace*{.2in}
Department of Physics (m/c 273), University of Illinois at Chicago,\\
\hspace*{.4in}845 W. Taylor Street, Chicago, Illinois 60607-7059 \\

\vspace*{1.0in}
\begin{abstract}
  Motivated by the concept of shape invariance in supersymmetric 
quantum mechanics, we obtain potentials whose spectrum consists 
of two shifted sets of equally spaced energy levels. These 
potentials are similar to the Calogero-Sutherland model except 
the singular term $\alpha x^{-2}$ always falls in the transition region
$-\frac{1}{4}<\alpha<\frac{3}{4}$ and there is a $\delta$-function
singularity at $x=0$. 
\end{abstract}
\newpage 
It is well-known
that a set of equally spaced energy eigenvalues corresponds 
to a particle moving in a harmonic oscillator potential. More specifically, 
in units of $\hbar = 2m = 1$, the set of energy levels 
\begin{equation} \label{en}
E_n = n \omega~~~~(n=0,1,2,...)
\end{equation}
corresponds to the one dimensional potential
\begin{equation} \label{vx}
V(x)=\frac{1}{4} \omega^2 x^2 - \frac{1}{2} 
\omega~~~~~(-\infty < x < \infty)~.
\end{equation}
However, if the particle is not allowed to move along the entire $x$-axis 
and is constrained to remain on the half-axis, a simple potential with 
eigenvalues given by eq. (\ref{en}) is 
\begin{equation} \label{vr}
V(r)=   \frac{1}{16} ~\omega^2 r^2 +\frac{l(l+1)}{r^2}- 
	\frac{\omega}{2}\left( l+ \frac{3}{2} \right) 
	~~~~(0 < r <\infty~;~~ l \ge -\frac{1}{2} )~.
\end{equation}
These potentials are not unique. It is well-known that they can be 
systematically deformed to produce multi-parameter isospectral families 
either via inverse scattering methods \cite{Chadan77} or using 
supersymmetric quantum mechanics techniques \cite{Cooper95}.

Now, instead of a single set of equally spaced eigenvalues, let us 
make an innocuous extension and ask for potentials which possess two 
shifted sets of equally spaced eigenvalues. The energy levels are 
given by $n(\omega_0 + \omega_1)$ and $n(\omega_0 + \omega_1)+\omega_0$ with 
$n=0,1,2,\cdots$. Clearly, there are two sets of energy levels with equal 
spacing $(\omega_0 + \omega_1)$, with one set displaced by $\omega_0$ 
with respect to the other set [see Fig. 1]. To our 
knowledge, potentials which possess the spectrum of Fig. 1 
have not been discussed in the 
literature.  We will obtain simple, explicit potentials using an 
approach motivated by recent developments \cite{Khare93,Barclay93} 
in obtaining new shape invariant potentials \cite{Gendenshtein83} in 
supersymmetric quantum mechanics \cite{Cooper95,Witten81}. These advances 
have been made using
novel choices of the function $f$ appearing in the change of 
parameters $a_1=f(a_0)$ in the shape invariance condition. 
In this paper, we use a transformation such 
that $f^2(a_0)=a_0$. The resulting shape invariant potentials are found to 
be singular but analytically solvable and will have a spectrum 
consisting of two shifted sets of equally spaced energy levels. 
Other potentials with the same spectrum can be readily obtained by making
isospectral deformations \cite{Cooper95,Khare89}, but this aspect will not 
be further developed here.
We discuss applications to multiparticle problems \cite{Calogero,Sutherland}. 
In particular, we find that there exist new eigenstates interlaced between 
the usual equally spaced Calogero spectrum of eigenvalues. These new 
states are a direct consequence of the fact that potentials with 
a $\alpha x^{-2}$ singularity with strength $\alpha$ in the transition 
region 
$-\frac{1}{4}<\alpha<\frac{3}{4}$, possess additional bound states with 
singular but normalizable wave functions.\sss

\noindent {\bf Shape Invariant Potentials:}\sms

Let us recall that supersymmetric partner potentials $V_{\pm}(x,a_0)$ are 
related to the superpotential $W(x,a_0)$ by 
\begin{equation} \label{vpm}
V_{\pm}(x,a_0)=W^2(x,a_0) \pm W'(x,a_0)~~,
\end{equation}
where $a_0$ is a set of parameters. These partner potentials are 
shape invariant if they both have the same $x$-dependence upto a change 
of parameters $a_1=f(a_0)$ and an additive constant $R(a_0)$. The 
shape invariance condition is
\begin{equation} \label{sipv}
V_+(x,a_0) = V_-(x,a_1) + R(a_0)~,
\end{equation}
or equivalently
\begin{equation} \label{sipw}
W^2(x,a_0)+W'(x,a_0) = W^2(x,a_1)-W'(x,a_1)+R(a_0)~~.
\end{equation}
The property of shape invariance permits an immediate analytic 
determination of energy eigenvalues \cite{Gendenshtein83}, eigenfunctions 
\cite{Dutt86} and scattering matrices \cite{Khare88}. For unbroken 
supersymmetry, the eigenstates of the potential $V_-(x)$ are:
$$
E_0^{(-)} =0~,~E_n^{(-)}=\sum_{k=0}^{n-1} R(a_k)~,
$$
\begin{equation}
\psi_0^{(-)} \propto e^{- \int^x_{x_0} W(y,a_0) dy}~,~
\psi_n^{(-)}(x,a_0)=\left[-\frac{d}{dx}+W(x,a_0)\right]
\psi_{n-1}^{(-)}(x,a_1)~~(n=1,2,3,...)~.
\label{algebra}
\end{equation}
Note the existence of a zero energy ground state, characteristic 
of unbroken supersymmetry.\sss

\noindent {\bf Change of Parameters:}\sms

Until 1992, the only known solutions of the shape invariance condition 
corresponded to a translational change of parameters
$$
a_1=f(a_0)=a_0+\beta~,~a_n=f^n(a_0)=a_0+n \beta~.
$$
Subsequently, the scaling change of parameters
$$
a_1=f(a_0)=q a_0~,~a_n=f^n(a_0)=q^n a_0
$$
was studied and shown to produce new classes of shape invariant 
potentials \cite{Khare93,Barclay93}, which included self-similar 
potentials \cite{Shabat92} 
as a special case. The above mentioned change of parameters 
are shown in Fig. 2. In general, the changes are monotonic, 
and the parameters 
never acquire the same values again. The only exceptions are 
the special cases 
$\beta=0$ and $q=1$, both of which have $f(a_0)=a_0$ and hence 
$a_n=a_{n-1}=\cdots=a_1=a_0$. For this situation, the energy 
levels are equally spaced 
since $E_n^{(-)}-E_{n-1}^{(-)}=R(a_0)$. If one denotes the constant value of 
$R(a_0)$ by $\omega$, then the shape invariance condition [eq. (\ref{sipw})] 
reads $W'(x,a_0)=\omega/2$. This gives $W(x,a_0)=\omega x/2$ and 
the standard harmonic oscillator potential $V_-(x)$ of eq. (\ref{vx}).

Let us now consider a new change of parameters defined by $f^2(a_0)=a_0$. 
Consequently,
$$
a_0=a_2=a_4=\cdots~~, ~~~f(a_0)=a_1=a_3=a_5=\cdots~~.
$$
There are many functions which meet the above requirement. 
One such function is $f(a_0)=C/a_0$, where $C$ is any real constant. 
Another choice is $f(a_0)=\frac{1}{2C} \log [\coth (C a_0)]$.
The energy levels will be 
spaced alternately by $R(a_0) \equiv \omega_0$ and 
$R(a_1) \equiv \omega_1$. This spectrum corresponds to two shifted sets 
of equally spaced eigenvalues.\sms

\noindent {\bf Single Particle Potentials:}\sms

To explicitly compute the 
potentials associated with two shifted sets of equally spaced eigenvalues, 
one needs to solve the shape invariance 
conditions:
$$
W^2(x,a_0)+W'(x,a_0) = W^2(x,a_1)-W'(x,a_1)+\omega_0~~,
$$
\begin{equation} \label{eq9}
W^2(x,a_1)+W'(x,a_1) = W^2(x,a_0)-W'(x,a_0)+\omega_1~~.
\end{equation}
The solution is readily obtained by straightforward manipulations.
The superpotential $W(x,a_0)$ is given by 
\begin{equation}
W(x,a_0)=\frac{(\omega_0+\omega_1)x}{4}+\frac{(\omega_0-\omega_1)}
{2x (\omega_0+\omega_1)}~,
\end{equation}
and the corresponding potential $V_-(x)$ is 
\begin{equation}
V_-(x,a_0)=
\frac{(\omega_0-\omega_1)(3\omega_0+\omega_1)}
{4{(\omega_0+\omega_1)}^2 x^2} 
- \frac{\omega_1}{2}
+\frac{{(\omega_0+\omega_1)}^2 x^2}{16}~.
\label{potential}
\end{equation}

The above results are correct everywhere except at the point $x=0$, 
where the superpotential $W(x,a_0)$ has an infinite discontinuity. 
Such a discontinuity is not acceptable, since while writing 
eqs. (\ref{vpm}) to (\ref{eq9}), we have implicitly assumed the existence 
of a continuous superpotential with a well-defined derivative giving 
rise to unbroken 
supersymmetry and a zero energy ground state. 
To study the problem further, consider a regularised, continuous 
superpotential $\tilde {W}(x,a_0,\epsilon)$ which 
reduces to $W(x,a_0)$ in the limit $\epsilon\to 0$. One such choice 
is
\begin{equation} \label{Wt}
\tilde {W}(x,a_0,\epsilon) = W(x,a_0)~f(x,\epsilon)
\end{equation}
where
\begin{equation}
f(x,\epsilon)= \tanh ^2 {\frac{x}{\epsilon}}~~.
\end{equation}
The moderating factor $f$ provides a smooth interpolation through the 
discontinuity, since it is unity everywhere except in a small region 
of order $\epsilon$ around $x=0$. In this region, 
$\tilde {W}(x,a_0,\epsilon)$ is linear with a slope 
$\frac{(\omega_0-\omega_1)} {2 \epsilon^2 (\omega_0+\omega_1)}$ 
which becomes larger as $\epsilon$ gets smaller. 
The potential $\tilde {V}_-(x,a_0,\epsilon)$
corresponding to the superpotential $\tilde {W}(x,a_0,\epsilon)$ is
\begin{equation} \label{Vte}
\tilde {V}_-(x,a_0,\epsilon) = \tilde {W}^2(x,a_0,\epsilon)-
\tilde{W}'(x,a_0,\epsilon)~.
\end{equation}
In the limit $\epsilon\to 0$, $\tilde {V}_-(x,a_0,\epsilon)$ reduces to
\begin{equation} \label{Vt}
\tilde{V}_-(x,a_0)=
V_-(x,a_0)-4~{W}(x,a_0)~\left[2 \theta(x)-1 \right]
~\delta(x)~,
\end{equation}
where $\theta(x)$ is the unit step function and we have used 
${\lim}_{\epsilon\to 0} 
~\frac{1}{2\epsilon} \sech^2\frac{x}{\epsilon} =\delta(x)$,
${\lim}_{\epsilon\to 0} 
~\tanh\frac{x}{\epsilon}= \left[2 \theta(x)-1 \right]$.
Thus we see that the potential $\tilde {V}_-(x,a_0)$ has an 
an additional singularity at the origin over ${V}_-(x,a_0)$ given by 
$\Omega(x) \equiv - \frac{2(\omega_0-\omega_1)} { x(\omega_0+\omega_1)}
~\left[2 \theta(x)-1 \right]~\delta(x)$. 
Under scaling this singularity transforms as 
$\Omega(\lambda x) = {1\over \lambda^2}\Omega(x) ~~~(\lambda \ge 0)$. The 
sign of this 
singularity  depends upon the values of $\omega_0$ and $\omega_1$.
It is best to consider the two cases $\omega_0>\omega_1$ and 
$\omega_0<\omega_1$ separately.\\
{\bf Case 1: $\omega_0>\omega_1$.}  An example of a superpotential for this 
case is shown in Fig. 3(a). The corresponding potential is drawn in Fig. 1(a), 
along with a few low lying energy levels. Note the sharp attractive shape of
the potential near $x=0$. In the limit $\epsilon\to 0$, 
this attractive $\delta$-function singularity at the 
origin is instrumental in producing a bound state at $E_0=0$.\\
{\bf Case 2: $\omega_0<\omega_1$.}  An example of a superpotential for this 
case is shown in Fig. 3(b). Again, proceeding as in case 1, one gets a 
repulsive singularity in the potential at the origin. This is shown in 
Fig. 1(b).

   Naively, in the limit $\epsilon \to 0$, the potential of eq. 
(\ref{potential}) appears identical to a three dimensional oscillator 
with a frequency $\omega \equiv \frac{(\omega_0+\omega_1)}{2}$ and angular 
momentum $l$ given by $l(l+1)=\frac{(\omega_0-\omega_1)(3\omega_0+\omega_1)}
{4{(\omega_0+\omega_1)}^2}$. However, there are some subtle but 
important differences. 
It is defined over the entire real axis ($-\infty<x<\infty$) and not just the
half line, and a very interesting feature of this
potential is the inevitable ``softness" of the inverse square term. For all 
values of parameters $\omega_0$ and $\omega_1$, the coefficient of the 
inverse square term is bounded between $-{1\over 4}$ and $3\over 4$; thus
the potential falls exactly in what is known as the transition
region\cite{Landau,Frank,Gangopadhyaya}. More specifically, for case 1 
$\left( \omega_0>\omega_1 \right)$, one has $0<l(l+1)<\frac{3}{4}$ and 
for case 2 $\left( \omega_0<\omega_1 \right)$ one has $-\frac{1}{4}<l(l+1)<0$. 
The important special case of the one dimensional harmonic 
oscillator has $\omega_0=\omega_1$: it corresponds to 
$l(l+1)=0$ and no $x^{-2}$ singularity. 
For transition potentials, both 
solutions of the Schr\"odinger equation are square integrable at the origin.
If one seeks eigenstates in a semi-infinite domain 
($0\leq x<\infty$), one just keeps the less ``singular'' solution of the 
two\cite{Landau,Frank,Gangopadhyaya}. 
However, for a fully infinite domain ($-\infty<x<\infty$) with a 
soft singularity at the origin, both are acceptable square integrable 
solutions of the Schr\"odinger equation, and must be retained to form
a complete set. 
Single particle eigenstates for the potential $\tilde{V}_-(x,a_0)=V_-(x,a_0)+
\Omega(x)$, where $\Omega(x)$ represents the point singularity at the 
origin, are readily obtained from eq. (\ref{algebra}).
The lowest four are:
\sss
\begin{eqnarray}
\label{eigenstates}
&E_0=0;~~~~&
\psi_0 ~\propto ~x^{-{{\omega_0-\omega_1}\over 
{2(\omega_0+\omega_1)}}} ~
e^{-{1\over 8} (\omega_0+\omega_1)x^2}~,\\
&E_1=\omega_0;~~~&
\psi_1 ~\propto ~x^{1+{{\omega_0-\omega_1}\over {2(\omega_0+\omega_1)}}} ~
e^{-{1\over 8} (\omega_0+\omega_1)x^2}~,\nonumber \\
&E_2=\omega_0+\omega_1;~&
\psi_2 ~\propto ~
\left(   
{{\omega_0-\omega_1}\over {\omega_0+\omega_1}}-1 + 
\frac{\omega_0+\omega_1}{2} x^2 \right) 
x^{-{{\omega_0-\omega_1}\over {2(\omega_0+\omega_1)}}} 
~e^{-{1\over 8} (\omega_0+\omega_1)x^2}~, \nonumber \\
&E_3=2\omega_0+\omega_1;&
\psi_3 ~\propto ~
\left(   
-{{\omega_0-\omega_1}\over {\omega_0+\omega_1}}-3 + 
\frac{\omega_0+\omega_1}{2} x^2 \right) 
x^{1+{{\omega_0-\omega_1}\over {2(\omega_0+\omega_1)}}} 
~e^{-{1\over 8} (\omega_0+\omega_1)x^2}.\nonumber 
\end{eqnarray}
In Fig. 4, we have drawn the lowest few eigenfunctions 
for two choices of the parameters $\omega_0$ and 
$\omega_1$. Note that for $\omega_0>\omega_1$, the ground 
state eigenfunction diverges at the origin, whereas for 
$\omega_0<\omega_1$ it vanishes. For the intermediate situation 
$\omega_0 = \omega_1$, the ground state is finite at the 
origin and is just the standard Gaussian solution of a 
one dimensional oscillator. 
In Fig. 5, we have plotted the five lowest energy eigenvalues $E_0$ to $E_4$ 
as a function of 
$\alpha \equiv (\omega_0-\omega_1)(3 
\omega_0+\omega_1)/{4 (\omega_0+\omega_1)^2}$ 
holding $\omega \equiv (\omega_0+\omega_1)/2$ 
fixed. $\alpha$ is the coefficient of the 
singular term in eq. (\ref{potential}). $2 \omega$ is the equal energy 
spacing in the two sets of energy levels. General expressions for 
these eigenfunctions and corresponding eigenenergies are given by
\begin{eqnarray}
E_{2n}=n(\omega_0+\omega_1)~,~
~~~~~&&
\psi_{2n} \propto 
x^{-{d\over 2}} e^{-{1\over 4}\omega x^2} L_n^{-{d\over 2}-{1\over 2}}
\left[\frac{\omega  x^2}{2}\right]~, \nonumber \\
E_{2n+1}=n(\omega_0+\omega_1)+\omega_0~,~
&&
\psi_{2n+1} \propto 
x^{1+{d\over 2}} e^{-{1\over 4}\omega x^2} L_n^{{d\over 2}+{1\over 2}}
\left[\frac{\omega x^2}{2}\right]
\label{spectrum-2}
,
\end{eqnarray}
\sss
\noindent where $d \equiv {{\omega_0-\omega_1}\over {(\omega_0+\omega_1)}}$ ,
$\omega \equiv \frac{(\omega_0+\omega_1)}{2}$, and $L_n$ are the standard 
Laguerre polynomials.\sms

\noindent {\bf Two Particle Case:}\sms

Consider the case of two particles interacting via the potential 
$\tilde{V}_-(\frac{x_1-x_2}{\sqrt{2}},a_0)$
of eq. (\ref{Vt}), consisting of the Calogero part
${V}_-(\frac{x_1-x_2}{\sqrt{2}},a_0)$ and the additional 
$\delta$-function singularity. The Calogero part 
[eq. (\ref{potential})] has been extensively discussed 
in the context of solvable many body problems in one dimension \cite{Calogero}. 
However, as we mentioned earlier, if the coefficient $\alpha$
of the singular term in 
eq. (\ref{potential}) is chosen to be between $-{1\over 4}$ and $3\over 4$,
the ``softness" of the singularity permits the 
penetration of particles through the barrier at the origin. For the two 
particle case this means that particles are able to interchange their 
positions by passing through each other. This leads to the appearance of 
additional eigenstates that are absent for a potential with a hard singularity 
$(\alpha >{3\over 4})$, where only those wave functions that vanish at the 
origin are allowed. As a consequence of this 
``tunneling", eigenstates break into groups of two, with the splitting of 
the two states vanishing as the coefficient approaches $3\over 4$. 
However, in our case, for the potential 
$\tilde{V}_-(\frac{x_1-x_2}{\sqrt{2}},a_0)$,
particles also see the point singularity $\Omega(\frac{x_1-x_2}{\sqrt{2}})$ 
as they pass 
through each other. The corresponding Schr\"odinger equation is
\begin{equation}
\left\{ -\frac{\partial^2}{\partial x_1^2} 
-\frac{\partial^2}{\partial x_2^2} + 
{{\omega^2}\over 8} \left( x_1-x_2 \right)^2 + 
{{2 \alpha } \over {\left( x_1-x_2 \right)^{2} }   } 
+\Omega\left(\frac{x_1-x_2}{\sqrt{2}}\right)\right\}
\psi(x_1,x_2)=E\psi(x_1,x_2).
\label{Calogero-2}
\end{equation}
Changing variables to center of mass coordinate $R=\frac{x_1+x_2}{2}$, and
relative coordinate $x=\frac{x_1-x_2}{\sqrt{2}}$, and then eliminating the 
motion of the center of mass, we arrive at
\begin{equation}
\left\{ -\frac{d^2}{d x^2} + 
{{1}\over 4} \omega^2 x^2 + 
{\alpha \over{x^{2}}} + \Omega(x) \right\}
\psi(x)=E\psi(x).
\label{Calogero-2a}
\end{equation}
The domain of $x$ is the entire real line. Thus, as discussed earlier for 
the single particle case, both 
solutions of the Schr\"odinger equation are necessary 
to generate a complete set of 
eigenstates on the full line. Now with the identification 
$\omega_0=2 a \omega$ and $\omega_1=2 (1-a) \omega$, where 
$a=\frac{1}{2} \sqrt{1+4\alpha}$, eq. (\ref{Calogero-2a}) reduces 
essentially to eq. (\ref{vr}). From eq. (\ref{spectrum-2}), the energy 
eigenvalues are given by $$E_{2n}=2n\omega,~~E_{2n+1}=2n\omega + 2a\omega~.$$

\newpage
\noindent {\bf Three Particle Case:}\sms

The Schr\"odinger equation for a three particle system that we consider here 
is:
\begin{equation}
\left\{ -\sum_{i=1}^3  \frac{\partial^2}{\partial x_i^2} 
+ \sum_{\stackrel{i,j=1}{i\neq j}}^3 \left(
{{\omega^2}\over {8}} \left( x_i-x_j \right)^2 
+ {2 \alpha\over {\left( x_i-x_j \right)^{2} }}
+\Omega\left(\frac{x_i-x_j}{\sqrt{2}}\right) \right)
\right\}\psi=E\psi~,
\end{equation}
where the singular function $\Omega\left( x \right) $ is given by 
$\Omega\left( x \right) = -\frac{4 b\left[2 \theta(x)-1 \right]}{x} 
~\delta(x)$, where $b$ is defined by $\alpha=b(b+1)$.
To avoid ``falling to the center''\cite{Landau}, we shall confine 
our attention to $\alpha >-{1\over 4}$. Transforming to center of mass 
and relative coordinates
\begin{equation}
R=\frac{x_1+x_2+x_3}{3}~,~
x=\frac{x_1-x_2}{\sqrt{2}}~,~
y=\frac{x_1+x_2-2x_3}{\sqrt{6}},
\end{equation}
and then eliminating the center of mass variable, we get 
\begin{eqnarray}
\left\{   - \frac{\partial^2}{\partial x^2} 
	  - \frac{\partial^2}{\partial y^2}  
	  + {3\over 8} {\omega^2} \left( x^2+y^2 \right) 
	  +  \alpha \left( 
	     \frac{1}{x^2} 
	  +  \frac{4}{(\sqrt{3}~y-x)^2} 
	  +  \frac{4}{(\sqrt{3}~y+x)^2} \right) 
\right.  &&\\   
	   \left.
	  +\Omega\left( x \right) 
	  +\Omega\left(\frac{\sqrt{3}~y-x}{2}\right) 
	  +\Omega\left(\frac{\sqrt{3}~y+x}{2}\right) 
	  \right\}\psi    &=& E\psi. \nonumber
\end{eqnarray}
Variables $x$ and $y$ span a two dimensional configurational plane. 
We parametrize this plane in polar coordinates $r=\sqrt{x^2+y^2}$ 
and $\phi=\tan^{-1}{x\over y}$. The range of these new variables are given by 
$0< r<\infty$ and $0\leq \phi <2\pi$. The Schr\"odinger equation in the 
variables $r$ and $\phi$ is 
\begin{eqnarray}
	  \left\{ 
	  - \frac{\partial^2}{\partial r^2} - {1 \over r}
	  \frac{\partial}{\partial r}  
	  +{3\over 8} {\omega^2} r^2 + 
	   {1 \over r^2} 
	   \left\{ 
	  -\frac{\partial^2}{\partial \phi^2} +  
\right.\right. && \\
\left.\left.
\sum_{n=0}^{2} 
	  \left(\frac{\alpha}{\sin^2\left(\phi+{{2n}\over 3} \pi\right) } 
	  +\Omega\left(\sin \left(\phi+{{2n}\over 3} \pi \right)  \right) 
	   \right) \right\}  \right\}  \psi
	  &=& E\psi; \nonumber
\end{eqnarray}
where we have used $\Omega\left(r \sin(\phi+{{2n}\over 3} \pi)  \right) 
=\frac{1}{r^2} \Omega\left(\sin(\phi+{{2n}\over 3} \pi)  \right) $. 
Owing to the zeros of the function $\sin(\phi+{{2n}\over 3} \pi)$, the 
$\Omega$-singularities are at $\frac{k\pi}{3} ~~~(k=0,1,\cdots,5)$ and they
can be written as $\sum_{k=0}^5\Omega(\phi-\frac{k\pi}{3})$.
Substituting $\psi(r,\phi)=R(r) f(\phi)$, and then using the usual procedure 
for separation of variables, we get:
\begin{equation}
\left( - \frac{\partial^2}{\partial r^2} - {1 \over r}
	  \frac{\partial}{\partial r}  
	  +{3\over 8} {\omega^2} r^2 + 
	   {b_l^2 \over r^2}  \right) 
	  R(r) = E R(r), 
\label{Calogero-3a}
\end{equation}
and
\begin{eqnarray}
\label{Calogero-3b}
\left\{
	  -  \frac{\partial^2}{\partial \phi^2} 
	  +   \alpha \left( 
	     \frac{1}{\sin^2\phi} 
	  +  \frac{1}{{\sin^2(\phi+{2\over 3} \pi) } }    
	  +  \frac{1}{{\sin^2(\phi+{4\over 3} \pi) } }    \right) 
\right. && \\ \left.
	  +  \sum_{n=0}^{n=2} \Omega\left({{\sin(\phi+{{2n}\over 3} \pi) } } 
	     \right) \right\} f_l(\phi) &=& b_l^2 ~f_l(\phi)~. \nonumber
\end{eqnarray}
The radial equation is same as that of a three dimensional oscillator. Its
eigenfunctions and eigenvalues are given by 
\begin{equation}
R_{nl}(r)=r^{b_l} {\rm exp}
\left[-{1\over 4} \sqrt{3\over 2}\omega r^2\right]
L^{b_l}_n\left[ \sqrt{3\over 8}\omega r^2\right],
 ~ ~ ~ ~ ~ ~ ~ ~ ~ ~ ~ 
E_n=\sqrt{\frac{3}{2}}\,(2n+b_l+1), \omega
\end{equation}
where $L^b_n$ are Laguerre polynomials. To determine energy eigenvalues 
$E_n$, we still
have to solve eq. (\ref{Calogero-3b}).
We first make a simplification using the identity \cite{Calogero}
$$\left( \frac{1}{\sin^2\phi} +  \frac{1}{{\sin^2(\phi+{2\over 3} \pi) } }    
+  \frac{1}{{\sin^2(\phi+{4\over 3} \pi) } }  \right) =
\frac{9}{{\sin^2(3\phi) } }.$$
Thus eq. (\ref{Calogero-3b}) now reduces to 
\begin{equation}
\label{Calogero-3c}
\left\{
	  - \frac{\partial^2}{\partial \phi^2} 
	  + \frac{9{\alpha}}{{\sin^2 3\phi } }    
	  + 9 \sum_{n=0}^{n=5} 
	    \Omega\left( 3\phi - n \pi \right) 
\right\}
	     f_l(\phi) = b_l^2 ~f_l(\phi) ~~,
\end{equation}
where once again we have used the scaling property of the singular function 
$\Omega$. 
Since the effective potential of eq. (\ref{Calogero-3c}) is periodic with a 
period $\pi/3$, we only need to obtain a solution in the domain 
$0<\phi<{\pi \over 3}$. Using even and oddness of eigenfunctions, 
solutions can be extended beyond $\phi=\pi/3$, as lucidly discussed 
in ref. \cite{Calogero}. 
Changing variables to $\xi=3\phi$ yields
\begin{equation}
\left\{
-\frac{\partial^2}{\partial \xi^2} 
+\frac{\alpha}{\sin^2\xi} 
+ \Omega\left( \xi \right) +  \Omega\left( \xi-\pi \right) 
\right\}
f_l(\xi) = B_l^2 ~f_l(\xi),
~
{\rm where}~~ B_l=\frac{b_l}{3}~~.
\label{Calogero-3d}
\end{equation}
The new domain is $0\leq \xi<\pi$. 
The major role played by the $\Omega$-singularity at the origin 
is to properly describe the discontinuity suffered by derivatives of the 
all even-states at $\xi=0$ and $\xi=\pi$.
Now we perform another change 
of the independent variable from $\xi$ to $z=\cos^2{\xi \over 2}$,
and define $f_l(z)=[z(1-z)]^\delta ~ h(z)$, where $\delta$ will be 
chosen judiciously to remove the singularity in eq. (\ref{Calogero-3d}).  
For $\delta=\frac{1 + \sqrt{1+4\alpha}}{4}$, we get
\begin{equation} \label{hyp}
z(1-z) \frac{d^2h}{dz^2} + 
\left[  2\delta + {1\over 2} - (1+4\delta) z \right] \frac{dh}{dz} - 
\left[4\delta^2-B_l^2 \right] ~ h(z)=0.
\end{equation}
This has the form of a hypergeometric equation and the general solution is
\begin{eqnarray}
f_l(z) &=&
c_1 z^{\delta} (1-z)^{\delta}
F\left[2\delta+B_l,2\delta-B_l,\frac{1}{2}+2\delta,z\right] \nn\\
&+& 
c_2 z^{\frac{1}{2}-\delta} (1-z)^{\delta}
F\left[\frac{1}{2}+B_l,\frac{1}{2}-B_l,\frac{3}{2}-2\delta,z\right] .
\end{eqnarray}
Usually one of these solutions has an unacceptable singularity near 
$z \approx 0$, and the other solution is retained. However, for
softly singular cases (-$\frac{1}{4}<\alpha<\frac{3}{4}$) both solutions are 
normalizable, and necessary to generate a complete set of eigenfunctions.
We would also like to emphasize that for the $\alpha \to 0$ limit, 
there is a 
smooth and continuous transition to one dimensional harmonic oscillator 
solutions (even and odd). The general solutions of eq. (\ref{Calogero-3c}) 
in the domain $0<\phi<\pi/3$ are given by
\begin{eqnarray}
f_{2l}(\phi) \sim 
\left(\cos{3\phi\over 2}\right)^{\frac{1}{2}-a}
\left(\sin{3\phi\over 2}\right)^{\frac{1}{2}+a} 
F\left[ -2l, 2l+1, 1-a; \cos^2{3\phi\over 2} \right]~,
&& b_{2l}=3 (2l+\frac{1}{2})~, \\
f_{2l+1}(\phi) \sim 
\left(\sin{3\phi}\right)^{\frac{1}{2}+a} 
F\left[ -2l-1, 2l+2+2a, 1+a; \cos^2{3\phi\over 2} \right]~,
&& b_{2l+1}=3 (2l+a+\frac{3}{2})~, \nonumber
\label{3-body}
\end{eqnarray}
where $a=\frac{\sqrt{1+4\alpha}}{2}$. 
The states $f_{2l}(\phi)$ ($f_{2l+1}(\phi)$) are symmetric 
(anti-symmetric) about the point $\phi={\pi \over 6}$. To determine wave
functions for any value of $\phi$, one uses 
$f_l(\phi)=f_l(\phi-\frac{m\pi}{3})$, where $m$ is an appropriately chosen 
integer such that $0\leq \phi-\frac{m\pi}{3}\leq \frac{\pi}{3}$.
The eigenvalues are given by 
$E_{n,2l} = \sqrt{\frac{3}{2}}\,(2n+6l+\frac{5}{2}) \omega$ and 
$E_{n,2l+1} = \sqrt{\frac{3}{2}}\,(2n+6l+3a+\frac{11}{2}) \omega$.

\sss
\noindent {\bf Discussion and Comments:}\sms

In this paper, we have shown that with a change of parameters of the type 
$a_2 \equiv f^2(a_0) = a_0$, one gets interesting shape invariant potentials 
with two shifted sets of equally spaced eigenvalues. These potentials are 
similar to the ones discussed by 
Calogero\cite{Calogero}. The main differences are: (a) the coefficient $\alpha$ 
of the inverse square term is always of the ``soft" type i.e. 
$-\frac{1}{4}<\alpha<\frac{3}{4}$ and (b) there is an additional 
$\delta$-function singularity at the origin (in the 
multi-particle case this is a 
two-body contact interaction). For multi-particle systems 
that interact via two-body potentials of the above type, the spectrum
includes the entire set eigenvalues of the Calogero potential as well as
another set of equally spaced eigenvalues which are slightly shifted from 
Calogero set.

Can the same technique 
be extended, for example, to $a_3 \equiv f^3(a_0) = a_0$, thus giving three
shifted sets of equally spaced eigenvalues?  This does not seem to be 
the case. Not only are there no simple real solutions to the constraint 
$f^3(a_0) = a_0$, but the shape invariance condition equations analogous to 
eq. (\ref{eq9}) are not easily solvable. Other choices for the change of 
parameters $a_1=f(a_0)$ are currently being investigated.

\sss

A.G. acknowledges the hospitality of the UIC Department of Physics where
part of this work was done. We also would like to thank R. Dutt and 
A. Khare for many valuable discussions and comments. 
Partial financial support from the U.S. 
Department of Energy is gratefully acknowledged.

\newpage
\noindent
{\Large \bf Figure Captions}\sss

\noindent
{\bf Figure 1:}  The potentials $\tilde{V}_-(x,a_0,\epsilon)$ given by 
eq. (\ref{Vte}) for the value $\epsilon=0.1$ and for 
two choices of parameters:
(a) $\omega_0 = 1.7$, $\omega_1=0.3$ ; (b) $\omega_0 = 0.3$, 
$\omega_1=1.7~$.
The potentials correspond to the superpotentials of Fig. 1. The
eigenvalue spectrum, consisting of two shifted sets of equally 
spaced energy levels, is given by eq. (\ref{spectrum-2}) 
and corresponds to the limit $\epsilon \to 0$.
\sss

\noindent
{\bf Figure 2:}  Graphical representation of various modes of changing parameters 
$a_n=f^n(a_0)$ in the shape invariance condition. The variation of $a_n$ 
versus $n$ is shown by the symbol $+$ for translations $a_n=a_0+n\beta$ and by 
the symbol $*$ for scaling $a_n=q^n a_0$. An example of the new change of 
parameters discussed in this 
paper is $f(a_0)=\frac{1}{2C} \log [\coth (C a_0)]$, and the results 
are shown by the symbol o. The numerical choices for this figure are 
$a_0=0.8,~\beta=0.1,~q=0.9,~C=0.5~$.
\sss

\noindent
{\bf Figure 3:}  The superpotentials $\tilde {W}(x,a_0,\epsilon)$ given by 
eq. (\ref{Wt}). The curves 
are drawn for a small value $\epsilon=0.1$, in order to see the behavior 
near the origin. There are two choices of parameters:
(a) $\omega_0 = 1.7$, $\omega_1=0.3$ ; (b) $\omega_0 = 0.3$, $\omega_1=1.7~$.
\sss

\noindent
{\bf Figure 4:}  The low lying eigenfunctions given by eq. (\ref{eigenstates}) 
corresponding to the potentials of Fig. 1. 
\sss

\noindent
{\bf Figure 5:}  A plot of the five lowest energy eigenvalues $E_0$ to $E_4$ as 
a function of $\alpha \equiv (\omega_0-\omega_1)(3 \omega_0+\omega_1)/{4
(\omega_0+\omega_1)^2}$ holding $\omega \equiv (\omega_0+\omega_1)/2$ 
fixed. $\alpha$ is the coefficient of the 
singular term in eq. (\ref{potential}). $2 \omega$ is the equal energy 
spacing in the two sets of energy levels. The two sets are displaced 
by energy $\omega_0$.
Regions shown by dashed lines 
correspond to singular but normalizable wave functions. 

\end{document}